\documentclass[sigconf,authorversion,nonacm]{acmart}

\usepackage{booktabs} % For formal tables
\usepackage{algorithm}
\usepackage{caption}
\usepackage{subcaption}
\usepackage[noend]{algpseudocode}
\usepackage{xcolor}
\usepackage{hyperref}
\usepackage{flushend}

 \setcopyright{none} 

\begin{document}
\title{A Framework for Efficient Memory Utilization in Online Conformance Checking}
% \titlenote{Produces the permission block, and
% \subtitle{Extended Abstract}
% \subtitlenote{The full version of the author's guide is available as
%   \texttt{acmart.pdf} document}
  
%\renewcommand{\shorttitle}{SIG Proceedings Paper in LaTeX Format}

\author{Rashid Zaman}
%\authornote{Corresponding author}
\orcid{0000-0003-2251-1090}
\affiliation{%
  \institution{Mathematics \& Computer Science\\
  Eindhoven University of Technology}
}
\email{r.zaman@tue.nl}

\author{Marwan Hassani}
\orcid{0000-0002-4027-4351}
\affiliation{%
 \institution{Mathematics \& Computer Science\\
  Eindhoven University of Technology}
}
\email{m.hassani@tue.nl}

\author{Boudewijn F. van Dongen}
  \orcid{0000-0002-3978-6464}
  \affiliation{%
 \institution{Mathematics \& Computer Science\\
  Eindhoven University of Technology}
  }
\email{b.f.v.dongen@tue.nl}

% \author{Rashid Zaman}
% \authornote{Corresponding author}
% \orcid{0000-0003-2251-1090}
% \affiliation{%
%   \institution{Eindhoven University of Technology}
%   \streetaddress{P.O. Box 513}
%   \city{Eindhoven} 
%   \state{The Netherlands} 
%   \postcode{5600 MB}
% }
% \email{r.zaman@tue.nl}

% \author{Marwan Hassani}
% \authornote{Corresponding author}
% % \authornote{The secretary disavows any knowledge of this author's actions.}
% \orcid{0000-0002-4027-4351}
% \affiliation{%
%   \institution{Eindhoven University of Technology}
%   \streetaddress{P.O. Box 513}
%   \city{Eindhoven} 
%   \state{The Netherlands} 
%   \postcode{5600 MB}
% }
% \email{m.hassani@tue.nl}

% \author{Boudewijn F. van Dongen}
% % \authornote{This author is the
% %   one who did all the really hard work.}
%   \orcid{0000-0002-3978-6464}
% \affiliation{%
%   \institution{Eindhoven University of Technology}
%   \streetaddress{P.O. Box 513}
%   \city{Eindhoven} 
%   \state{The Netherlands} 
%   \postcode{5600 MB}
% }
% \email{b.f.v.dongen@tue.nl}

% % The default list of authors is too long for headers}
% \renewcommand{\shortauthors}{R. Zaman et al.}

\begin{abstract}
% Process mining bridges the two important fields of data mining and process science. Process \textit{events} are of fundamental nature in process mining which represent the execution of activities in an individual instance or run of a business process called a \textit{case}. 
Conformance checking (CC) techniques of the process mining field gauge the conformance of the sequence of events in a case with respect to a business process model, which simply put is an amalgam of certain behavioral relations or rules. Online conformance checking (OCC) techniques are tailored for assessing such conformance on streaming events. The realistic assumption of having a finite memory for storing the streaming events has largely not been considered by the OCC techniques. We propose three incremental approaches to reduce the memory consumption in prefix-alignment-based OCC techniques along with ensuring that we incur a minimum loss of the conformance insights. Our first proposed approach bounds the number of maximum states that constitute a prefix-alignment to be retained by any case in memory. The second proposed approach bounds the number of cases that are allowed to retain more than a single state, referred to as multi-state cases. Building on top of the two proposed approaches, our third approach further bounds the number of maximum states that the multi-state cases can retain. All these approaches forget the states in excess to their defined limits and retain a meaningful summary of them. Computing prefix-alignments in the future is then resumed for such cases from the current position contained in the summary. We highlight the superiority of all proposed approaches compared to a state of the art prefix-alignment-based OCC technique through experiments using real-life event data under a streaming setting. Our approaches substantially reduce memory consumption by up to $80\%$ on average, while introducing a minor accuracy drop.  %The most commonly solution presented in the literature manages memory through forgetting the \textit{least recently updated} cases, assuming them to be closed with no further behavior expected to be observed for these cases. Such assumption may not always hold and subsequently future behavior may be observed for such forgotten cases to be termed as \textit{orphan events}. Without some treatment in lieu of their forgotten prefixes, these events will be improperly marked as non-conformant, the phenomenon to be referred to as \textit{missing-prefix problem}. 
 %Prefix-alignments consist of states which embed events and their conformance information. 
%For all the cases stored in memory, this approach keeps forgetting states in a first-in-first-out fashion with retaining only the specified number of latest states and a summary of the forgotten state's prefix.
 %The single state of the rest of the cases contains a summary of the forgotten states prefix. %To reduce the loss of conformance insights, this approach forgets the cases in excess to the defined limit through a systematic forgetting approach. Both these approaches retain some minimal information regarding the forgotten events, or states to be precise, to resume the conformance checking from the current position.  %While, the third class of \textbf{approaches} bounds the memory with both the number of states per case and the number of cases to be stored in memory simultaneously.
 %different type of information regarding the forgotten cases. On observing the orphan events for the forgotten cases, the proposed techniques enable the prefix alignments to be calculated from some reachable marking of the initial marking of the process model, with taking into consideration the retained information for these cases.
\end{abstract}


%\ccsdesc{Prefix-Alignments~Memory-Bounding}


\maketitle

\section{Introduction}

Process mining~\cite{van2016data,carmona2018conformance}, a young discipline, bridges the gap existing between data mining and process science~\cite{article} fields. Business \textit{process models} and \textit{event} data are its most prominent inputs. A process model depicts a business process. It can also be viewed as an amalgam of the business rules or behavioral relations which guide the execution sequence of the various activities in a business process. An event represents the execution of an atomic activity in a business process. A business process therefore serves as a \textit{concept} in the process mining landscape. Events belonging to the same instance or run of a process are treated as a single \textit{case}. A case syntactically acts as a data point for process mining. Events, as the entities constituting a case, therefore are analogous to the \textit{features} of a case. A case is usually expected to follow a sequence of the behavioral relations inscribed in the process model, implying that the order of the events is also important.

\begin{figure*}
\includegraphics{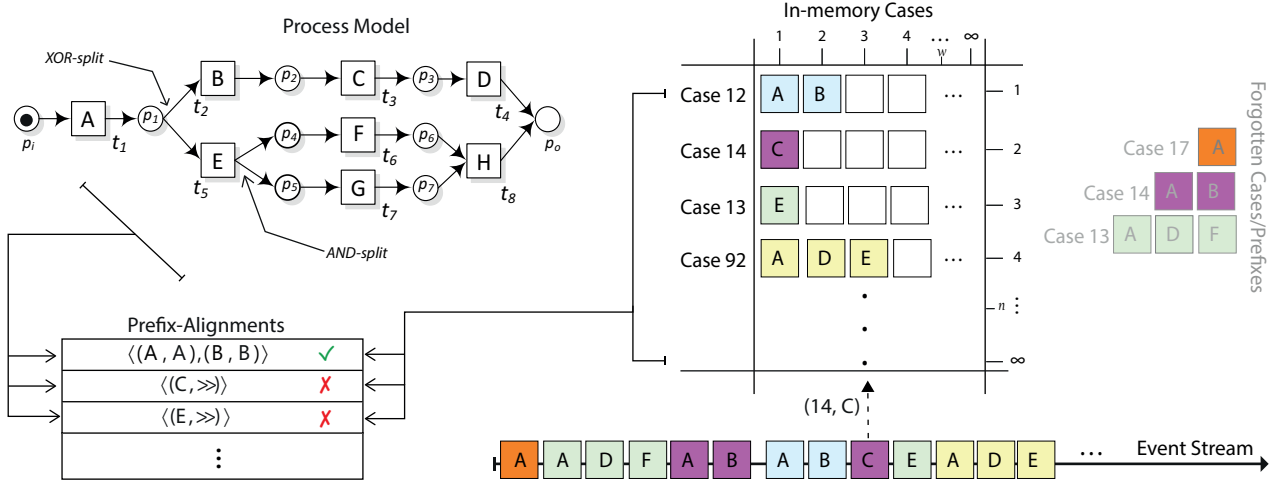}
\caption{A simple overview of an online conformance checking (OCC) scenario.}
\label{fig:overview}
\end{figure*}

Like data streams, event streams are characterized by a high throughput of events generated by a single or multiple concept instances running in parallel. Online process mining techniques~\cite{Burattin2018} are tailored to process such event streams. There, online conformance checking (OCC) techniques analyze in-progress cases for the purpose of checking their conformance to the relevant process model. CC techniques require the sequence of all the events constituting a case rather than individual events. This requirement implies that even previously processed events shall be retained in order to ensure the completeness of its relevant case. The restriction of having a finite memory for coping with infinite streams is an established concern for data streams processing~\cite{hassani2015efficient}. The aforementioned requirement makes the finite memory constraint much more involved for OCC techniques.

In order to reduce memory consumption, OCC literature~\cite{burattin2018online,van2019process} suggests forgetting the least recently updated cases, assuming them to be concluded. While this assumption may hold in some process domains, usually cases in real-life processes exhibit very diverse behavior with respect to the temporal distribution of their events. Additionally, the number of active in-parallel running cases may be higher than that allowed by the aforementioned techniques to be stored in the memory simultaneously. Accordingly, events belonging to the forgotten cases may be observed in the future, referred to as \textit{orphan events} in the literature~\cite{10.3389/fdata.2021.705243}. Such orphan events, without undergoing any remedial treatment, will be considered by the CC techniques to be belonging to newly initiated cases. Accordingly, these events will be improperly marked as nonconforming to the process model, primarily because of their forgotten prefix. This improper penalization is termed as \textit{missing-prefix} problem in the literature~\cite{10.3389/fdata.2021.705243}. 

Alignments-based CC techniques~\cite{adriansyah2014aligning} discover a sequence of activities in the relevant process model, referred to as an \textit{alignment}, which maximally matches the sequence of the activities represented by the events in a case. Alignments encode events, their counterpart activity in the reference process model, and their conformance statistics as \textit{moves}. Prefix-alignments~\cite{adriansyah2013controlling} is a variant of conventional alignments which can properly deal with incomplete cases. Moves are termed as \textit{states} by the incremental prefix-alignments~\cite{van2019online}. Figure~\ref{fig:overview} provides a simplistic overview of a prefix-alignment-based OCC setup. An event observed on the event stream is appended to its respective case in an infinite memory. This updated case is accordingly subjected to CC with the reference process model by the prefix-alignments. If we limit the memory through bounding the number of events in cases or bounding the number of maximum cases then we have to forget prefixes of cases or entire cases respectively. Orphan events for such partially or fully forgotten cases, for instance, Case ``14'', are improperly marked as non-conforming to the reference process model, as depicted in its prefix-alignment.

We present three incremental approaches to effectively reduce the memory footprint or in other words accommodate more cases using the same storage in prefix-alignments-based OCC techniques. Our proposed approaches partially or entirely forget the states constituting the prefix-alignments of cases and yet avoid the missing-prefix problem.

Our first proposed approach imposes a limit on the number of states to be retained by any case in the storage. In a first-in-first-out fashion, states in excess of the specified limit are forgotten. We avoid the missing-prefix problem by resuming the prefix-alignments computation for the orphan events from the position reached by the forgotten prefix states, through retaining their \textit{summary} as a special state. Our second approach imposes a limit on the number of multi-state cases, i.e., cases that can grow freely in terms of the states constituting their prefix-alignments. For this purpose, we prudently forget cases through a defined \textit{forgetting criteria} to conform to the defined limit. The main intuition behind our forgetting criteria is the maximization of the probability of correctly estimating the conformance of the cases with orphan events. To further reduce the memory consumption, our third proposed approach imposes a limit on the number of states to be retained by the fixed number of multi-state cases. As with the first approach, we retain a summary of the forgotten prefix states in the second and third approaches as well in order to avoid the missing-prefix problem.

Through experiments with real-life event data, by emitting its events on the basis of their actual timestamps as a stream, we demonstrate the effectiveness of our proposed approaches. We conclude that without any loss on the estimated (non)conformance in the first approach and with tolerable loss in the second and third approach, the memory footprint and the number of computations of the prefix-alignments-based OCC techniques can be significantly reduced through adopting these approaches.

This paper extends the approaches and the results presented in the short paper of the authors \cite{SAC2022} with two additional novel memory utilization methods and a further extensive experimental evaluation.

The rest of this paper is organized as follows. Section~\ref{sect:relatedwork} provides an overview of the existing relevant work. Section~\ref{sect:preliminaries} defines and explains some key concepts which are necessary for elaborating our three proposed approaches which we present in Section~\ref{sect:proposedtechniques}. Details and findings of the experiments conducted for evaluation of the proposed approaches are provided in Section~\ref{sect:evaluation}. Finally, Section~\ref{sect:conclusions} concludes the paper along with some ideas for future work.

\section{Related Work}\label{sect:relatedwork}

A landscape of the online process mining techniques was provided by \cite{Burattin2018}. Being part of the process mining manifesto~\cite{van2011process}, OCC is receiving attention from the research community. Using regions theory~\cite{van2007iterative}, \citep{burattin2017framework} added deviating paths to the process model and accordingly non-conformant cases following those paths are detected. Alignments~\citep{adriansyah2014aligning} is the state-of-the-art underlying technique for CC of completed cases~\cite{carmona2018conformance}. Accordingly, prefix-alignments~\cite{adriansyah2013controlling} were introduced for CC of incomplete cases with a process model, without penalizing them for their incompletion. Decomposition~\cite{van2013decomposing} and later Recomposition~\cite{lee2018recomposing} techniques were proposed to divide and then unite the alignments computation for efficiency. 

Incremental prefix-alignments~\citep{van2019online} combined prefix-alignments with a lightweight model semantics based method for efficiently checking conformance of in-progress cases in streaming environments. The technique of ~\citep{burattin2018online} determines the conformance of a pair of streaming events by comparing them to the behavioral patterns constituting the reference process model. Their technique is computationally less expensive in comparison to alignments and is also able to deal with a warm start scenario, where the event in an observed case does not represent a starting position. This approach somehow abstracts from the reference process model and markings there, therefore, it is hard to closely relate cases to the reference process model. The online conformance solution of \cite{LEE2021101674} used Hidden Markov Models (HMM) to first locate cases in their reference process model and then assess their conformance.

The availability of a limited memory, implying the inability to store the entire stream, has been sufficiently investigated in the data stream mining field~\citep{bahri2021data, gomes2019machine}. In contrast, the memory aspect of online process mining in general and OCC in specific has attracted less attention in the literature. The majority of such techniques therefore assume the availability of an infinite memory. \citep{hassani2019application} generically suggested maintaining an abstract intermediate representation of the stream to be used as input for various process discovery techniques. \citep{burattin2018online} limited the number of cases to be retained in memory by forgetting \textit{inactive} cases. Forgetting cases on inactivity criteria may lead to the missing-prefix problem in many process domains. Prefix imputation approach~\cite{10.3389/fdata.2021.705243} has been proposed as a two-step approach for bounding memory but at the same time avoiding missing-prefix problem in OCC. The technique selectively forgets cases from memory and then imputes their orphan events with a prefix guided by the normative process model.

\section{Preliminaries}\label{sect:preliminaries}

In this section, we briefly explain the concepts related to our proposed approaches. 

\paragraph{\textbf{Process model}:} A business process is the \textit{concept} in the process mining domain, which is represented by a \textit{process model}. A process generates the \textit{events} which constitute the data points, i.e., \textit{cases}, of process mining. Activities are spatially laid down in accordance with certain behavioral relations to synthesize a process model. For instance, activity \textit{A} in a process must always be followed by activity \textit{B}, the relation termed as a sequence. Other relations include choices, concurrency and looping. While multiple representations exist for modelling a process, we use the highly formal Petri nets. A \textit{Petri net} is represented as a tuple $N = (P, T, F, \lambda)$, where $P$ represents a finite set of places, $T$ a finite set of transitions, $F \subset(P \times T) \cup (T \times P)$ a set of flow relations between places and transitions. $\lambda$ is a labelling function assigning transitions in $T$ with labels from the set of the activity labels $\Lambda \in \mathcal{A}$, where $\mathcal{A}$ is the universe of activities.

The stage or position of a case is represented through a \textit{marking} $M$ in the process model. A marking $M$ is essentially a multiset of tokens over the places $P$ in the process model $N$, i.e., $M: P \rightarrow \mathbb{N}$. The \textit{initial Marking} $M_{i}$ is the stage which ideally every case shall start in and the \textit{final marking} $M_{f}$ is a stage where ideally every case shall eventually end. Apart from regular transitions (representing business activities), \textit{silent transitions} $\tau$ (Taus) are used in Petri nets for the completion of routing. A transition $t$ having a token in each of its input places, as part of a marking $M$, is said to be \textit{enabled}, represented as $M [ \, t \rangle$. An enabled transition can \textit{fire} or \textit{execute} thereby consuming a token from each of its input places and accordingly producing a token in each of its output places, resulting in a change of marking from $M$ to $M^\prime$ represented as $M [ \, t \rangle M^\prime$. The consecutive firing of a sequence of (enabled) transitions $\sigma\in T\mbox{*}$ starting from a marking $M$ and leading to $M^\prime$ is referred to as a \textit{firing} or \textit{execution sequence} of $M$ and represented as $M \xrightarrow{\sigma}M^\prime$. Typically, the set of execution sequences for a process model $N$ is finitely large in absence of loops and infinitely large in presence of a loop. An execution sequence $\sigma$ starting from $M_{i}$ and ending in $M_{f}$ is referred to as a \textit{complete execution sequence} of $N$, i.e., $M_{i} \xrightarrow{\mathit{\sigma}} M_{f}$.

The top left side of Figure \ref{fig:overview} contains an example process model as a Petri net. The circles $\{p_1, p_2, \dots\}$ represent places. Rectangle-shaped transitions $\{t_1, t_2, \dots\}$ map the process model to the corresponding process activities through labels $\{A, B, C, \dots\}$. Directed arcs $F$ through connecting places and transitions enforce the behavioral relations. For instance, the output arc from transition $t_2$ enters place $p_2$ which is connected to transition $t_3$ through its output arc, hence the two transitions form a \textit{sequential} relation. Transitions $\{t_2, t_3, t_4\}$ are in a \textit{choice} relation with transitions $\{t_5, t_6, t_7, t_8\}$ through an \textit{XOR-split}. Similarly, transitions $t_6$ and $t_7$ are \textit{parallel} by virtue of an \textit{AND-split}. With the start place $p_i$ having a single token, the Petri net is in initial marking $[{p_i}^1]$, or simply $[p_i]$. Only transition $t_1$ is enabled in this marking and its firing results in marking $[p_1]$. $\langle t_1, t_2, t_3 \rangle$ is one of the execution sequences, resulting in marking $[p_3]$ which is therefore a reachable marking of the $M_{i}$. The sequence $\langle t_1, t_2, t_3, t_4\rangle$ represents a complete execution sequence which puts the token in place $p_{o}$ indicating that final marking $M_{f}$ is reached. For sound understanding of the mentioned and other related concepts, interested readers are referred to~\citep{van2016data}.

\begin{table}
  \caption{An example event log.}
  
  \label{tab:eventlog}
  \begin{tabular}{ccccl}
    \toprule
    Event id & Case id & Activity & Timestamp\\
    \midrule
     1 & 1 & A & 2021-10-01 12:45 \\
     2 & 2 & A & 2021-10-01 13:03 \\
     3 & 1 & B & 2021-10-02 10:07 \\
    4 & 2 & B & 2021-10-09 14:31  \\
     5 & 3 & A & 2021-10-09 17:29  \\
    6 & 3 & E & 2021-10-13 16:49 \\
    7 & 3 & F & 2021-10-13 16:59 \\
    8 & 3 & G & 2021-10-20 11:23 \\
   9 & 1 & C & 2021-10-20 11:23 \\
    $\vdots$ & $\vdots$ & $\vdots$ & $\vdots$ \\
    \bottomrule
   
  \end{tabular}
\end{table}

\paragraph{\textbf{Events}:} The execution of activities in a business process are logged in the form of events. Events belonging to the same case are ordered to generate an \textit{event log} $L$. Formally, an event $e$ minimally consists of 1) the case identifier to which the event belongs, 2) the corresponding activity name represented as $\#_{activity}(e)$, and 3) the timestamp of the execution of the corresponding activity represented as $\#_{time}(e)$. It is important to note that every event is unique and distinct. Events referring to exactly the same activity, having the same timestamp, and belonging to the same case are by context different and distinct. The sequence of events $\sigma$ belonging to a case $c$ is referred to as the \textit{trace} of the case $c$. $|\sigma|$ denotes the length, i.e., the number of events, in a trace.

Table~\ref{tab:eventlog} depicts an excerpt of example event log $L$ obtained through firing of some execution sequences of the process model contained in Figure~\ref{fig:overview}. Each row in Table~\ref{tab:eventlog} represents an event $e$. For instance, the first row is event $e_1$ with event id of ``1'' and corresponds to execution of the activity $\#_{activity}(e)$ = \textit{A} in the context of Case ``1'' and at timestamp $\#_{time}(e)$ $ = ``2021-10-01$ $12:45$''. Events $e_5, e_6, e_7$ and $e_8$ constitute the trace for Case ``3'', which in essence corresponds to an execution sequence of the process model depicted in Figure~\ref{fig:overview}. The trace for Case ``3'' can \textit{simply} be denoted in the sequence of activities form as $\langle A, E, F, G\rangle$. Notice that all these cases are running in parallel. Event streams are characterized by a large number of in-parallel running cases.

\paragraph{\textbf{Events stream}:} An event log $L$ consists of historical or completed cases. On contrary, the cases observed on a stream evolve and new cases arrive as well. Formally, let $C$ be the universe of case identifiers and $\mathcal{A}$ be the universe of activities. An event stream $S$ is an infinite sequence of events over $C \times \mathcal{A}$, i.e., $S \in (C \times \mathcal{A})^*$. A stream event is represented as $(c, a) \in C \times \mathcal{A}$, denoting that activity $a$ has been executed in the context of case $c$. Like events in general, every stream event is unique and distinct, even if their respective activities, the case identifiers, and even the arrival times are exactly the same. Observed stream events are required to be stored under the notion of their respective cases. Event streams are characterized by continuous and unbounded emission of stream events $(c, a)$.

\paragraph{\textbf{Conformance Checking}:} By virtue of a variety of contextual factors~\cite{van2016data,carmona2018conformance}, activities in cases are executed in diverse ways. The sequence of these activities may even deviate from the behavioral relations envisaged as a process model. Conformance Checking (CC) is the comparison of cases with their reference process model to highlight deviations, if any. The detected deviations provide insights for remedial actions to mitigate the sources of non-conformance.

Many techniques have been developed but \textbf{alignments} have been positioned as the de facto standard technique for checking conformance of cases. Alignments, represented as $\gamma$, explain the sequence of events (or simply activities) in a case through a complete execution sequence in the reference process model~\citep{carmona2018conformance}, i.e., $M_{i} \xrightarrow{\mathit{\sigma}} M_{f}$. A case is considered conformant or fitting if a complete execution sequence of the process model exists that fully explains or reproduces its trace, otherwise non-conformant. The extent of the non-conformance of cases is measured through the degree of mismatch existing between their traces and a maximally-explaining complete execution sequence, termed as optimal alignment $\gamma_{opt}$.

\begin{figure}
\includegraphics{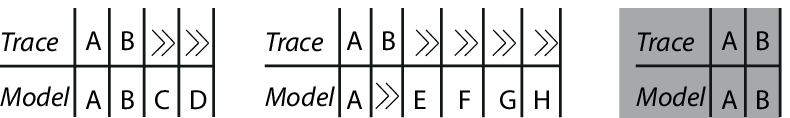}
\caption{Example (prefix)alignments for Case ``2'' of event log of Table~\ref{tab:eventlog} with the process model of Figure~\ref{fig:overview}.}
\label{fig:alignment}
\end{figure}

The non-shaded part of Figure~\ref{fig:alignment} shows two of the many possible alignments of the trace for Case ``2'', i.e., $\langle A, B\rangle$, of the event log depicted in Table~\ref{tab:eventlog} with the process model of Figure~\ref{fig:overview}. The \textit{trace} part of these alignments (neglecting $\gg$'s) is the same as the trace for Case ``2'' while their \textit{model} part (neglecting $\gg$'s) is a complete execution sequence of the process model. A corresponding trace and model entry in the alignment is known as a \textit{move} and represented as a pair, for instance, $(A, A)$. Moves without skip-symbol $\gg$ are termed as \textit{synchronous moves} and imply that an enabled transition with the same label as the activity represented by the event of the pair is available in the current marking. Moves with $\gg$ in the trace part of the pair are referred to as \textit{model moves} and illustrate that the trace is missing an activity for the transition in the pair which is enabled in the current marking and the execution sequence requires it to be fired. Moves with $\gg$ in the model part of the pair are referred to as \textit{log} moves which signal the missing of a fired transition in the complete execution sequence for the activity in the trace part. For instance, the first move in the middle alignment of Figure~\ref{fig:alignment} is a synchronous move, while the second and third moves are log and model moves respectively.

As evident from Figure~\ref{fig:alignment}, multiple alignments of a trace with a process model are possible. Therefore, moves are associated with a \textit{move cost} in order to rank alignments. Usually, synchronous moves and model moves with silent transitions $(\gg, \tau)$ are assigned a zero cost. The sum of the costs of all the individual moves of an alignment  is referred to as \textit{trace fitness cost} which is represented as $\kappa(\gamma)$. CC looks for an optimal alignment $\gamma_{opt}$ which bears the least trace fitness cost. It is worth mentioning that even multiple optimal alignments may exist for the same trace. 

Alignments assume cases to be completed while cases in an event stream may not be completed yet. For checking the conformance of such evolving cases, the \textbf{prefix-alignments} variant of the alignments is more appropriate. Prefix-alignments, represented as $\overline{\gamma}$, explain the sequence of events in a case through an execution sequence of the process model, rather than a complete execution sequence. The rest of the concepts, such as moves and their associated costs, are the same as for alignments.

Consider the trace for Case ``2'', i.e., $\langle A, B\rangle$, of the event log of Table~\ref{tab:eventlog}. The shaded part of Figure~\ref{fig:alignment} shows one of the possible prefix-alignments of this trace with the process model of Figure~\ref{fig:overview}. The \textit{trace} part of this prefix-alignment still corresponds to the trace while the \textit{model} part is an execution sequence of the process model. As with conventional alignments, different types of moves in prefix-alignments are associated with move costs in order to rank them for identifying the optimal one, i.e., $\overline{\gamma}_{opt}$. The optimal prefix-alignment evolves and changes with the evolution of the trace.

In contrast to a single alignment computation per case in conventional alignments, a prefix-alignment needs to be computed upon observing every single stream event $(c, a)$ in event streams. (Prefix)alignments computation through a shortest path search in a synchronous product~\cite{adriansyah2011cost} is quite compute-intensive. Therefore, the approach in~\cite{van2019online} tailored prefix-alignments to be efficient in checking conformance on streaming events, referred to as incremental prefix-alignments in this work. This approach first checks if activity $a$ of the observed stream event $(c, a)$ corresponds to a transition $t \in T$ that is enabled in the marking $M$ of the previously computed prefix-alignment $\overline{\gamma}$ of $c$ (or $M_{i}$ if the event is the first for the case). If found, a synchronous move $(a,t)$ is appended to the previously computed prefix-alignment existing in case-based memory $D_{\mathcal{C}}$. We refer to this method of extending a prefix-alignment as \textit{model semantics based prefix-alignment}. If unsuccessful, then a fresh optimal prefix-alignment $\overline{\gamma}_{opt}$ is computed for the trace of $c$ through a shortest path search in the synchronous product, starting from the initial marking $M_{i}$. We refer to this method as \textit{shortest path search based prefix-alignment}. The latter method is computation-wise expensive than the former. Moves are stored as \textit{states} constituting the prefix-alignments of cases. A prefix-alignment therefore can simply be represented as a sequence of states which we represent as ${\langle \overline{\gamma}(i) \rangle}_{i = 1 \dots z}$, where $i$ is the index of the state in the prefix-alignment $\overline{\gamma}$ and $z$ is the total number of states in $\overline{\gamma}$, i.e., $z=|\overline{\gamma}|$. A state $\overline{\gamma}(i)$ additionally stores its move cost and the marking $M$ of its case reached through the states \{$1,2, \dots , i-1, i$\}. We can represent this marking $M$ as ${M}(\overline{\gamma}(i))$. The term state may not be confused with that used in the process mining literature to represent a marking. Interested readers are referred to~\citep{carmona2018conformance} for a deeper understanding of the CC, alignments, prefix-alignments, and related concepts.

\section{Our Memory-Efficient Framework For Online Conformance Checking }\label{sect:proposedtechniques}
For activity $a$ of an observed stream event $(c,a)$, the model semantics based prefix-alignments method requires only the current marking $M$ of its previously computed $\overline{\gamma}$ to extend it for $a$. In contrast, the shortest path search based prefix-alignment reverts to $M_{i}$, but only for the sake of computing an optimal $\overline{\gamma}$ for the trace of $c$. Exploiting the aforementioned facts, we present three effective memory consumption approaches in the context of prefix-alignments-based OCC techniques.

\subsection{Bounded states with carryforward marking and cost}\label{sect:fixedstates}

As our first approach, we define a limit $w$ on the number of states that can be retained by the prefix-alignments $\overline{\gamma}$ of cases in $D_{\mathcal{C}}$, i.e., $\forall_{c \in D_{C}} |\overline{\gamma}_{c}|\leq w$. Upon observing an event for an existing case, after the computation of its prefix-alignment, we forget the earliest prefix state(s) in excess of the defined limit. However, we retain a \textit{summary} of the forgotten states as a special state prepended to the surviving states such that the total number of states is $w$. This summary consists of marking $M$ reached with the forgotten states and the cumulative cost of its moves $\kappa_{o}$. When a shortest path search based prefix-alignment is necessitated, we resume the prefix-alignment computation of the orphan events from $M$, i.e., provisionally ${M_{i}} = M$, thereby avoiding the missing-prefix problem. Similarly, we minimize the underestimation of their fitness costs by adding the cost incurred by the forgotten states as \textit{residual} to that incurred by the orphan events, i.e., $\kappa_{o} + \kappa({\overline{\gamma}}^\prime)$. The maximum memory required therefore is reduced to $|D_\mathcal{C}| \times w$ states. Algorithm~\ref{alg:state-based} provides an algorithmic summary of the proposed approach.

\begin{algorithm}[t]
\small
	\caption{Prefix-alignment-based OCC with bounded states}\label{alg:state-based}
	
	\begin{algorithmic}[1]
     \Require {$S \in (C \times \mathcal{A})^\ast, w$}
 	\State $i \leftarrow 0$
    \While{true}
		
			\State $i \leftarrow i+1$;
			\State $(c,a) \leftarrow S(i)$;
			\State $\overline{\gamma} \leftarrow D_{\mathcal{C}}(c,\;i-1)$;
			\State copy all alignments of $D_{\mathcal{C}}(c^\prime,\;i-1)$ to $D_{\mathcal{C}}(c^\prime,\;i)$ for all $c^\prime \in \mathcal{C}$;
 			  		\small
  		\State compute $\overline{\gamma}^\prime$ through \textit{model semantics} or \textit{shortest path search}~\cite{van2019online}
  		\If {$|\overline{\gamma}^\prime| > w$}
  		 \State ${\overline{\gamma}}_{o} = \langle (\emptyset,\kappa({\langle \overline{\gamma}^\prime(i) \rangle}_{i = 1 \dots |\overline{\gamma}^\prime|-w+1}), M({\overline{\gamma}^\prime}(|\overline{\gamma}^\prime|-w+1)))\rangle$ 
  		 \State $\overline{\gamma}^\prime = \overline{\gamma}_{o} \cdot {\langle \overline{\gamma}^\prime(i) \rangle}_{i = |\overline{\gamma}^\prime|-w+2 \dots |\overline{\gamma}^\prime|}$;
  		  \small
  		 \EndIf
  		 \State $D_{\mathcal{C}}(c,\;i) \leftarrow \overline{\gamma}^\prime$; 
  		\EndWhile 
	\end{algorithmic} 
\end{algorithm}

With forgetting states, we actually forget the embedded events as well. The model semantics based prefix alignment will not be affected by this forgetting of events as it requires only the current marking $M$ reached by the prefix alignment of the previously observed events, and not the events themselves. The shortest path search based prefix alignment however will revisit the prefix alignment computation from $M$ for only the events represented by the retained states. As a consequence, the freshly computed prefix alignment may not be a global optimum. This approach also reduces the number of computations performed in shortest path search based prefix alignment as the prefix-alignments are computed only for a subsequence of the observed events for cases whose prefix states have been forgotten.

\subsection{Bounded cases with carryforward marking and cost}\label{sect:fixedcases}

The approach presented in Section~\ref{sect:fixedstates} reduces the memory consumption to $|D_\mathcal{C}| \times w$ states but it can still grow unbounded after storing an infinite number of cases, i.e., when $|D_\mathcal{C}|\rightarrow \infty$. In order to further reduce the memory consumption by $D_\mathcal{C}$, we define a limit $n$ on the maximum number of multi-state cases. Hence, only $n$ cases can retain more than a single state in their prefix-alignments, i.e., $|\overline{\gamma}| \geq 1$. For all the other cases, their prefix states are forgotten and only a single special state is retained in a repository $R_{C}$. This single state contains the summary of the forgotten prefix states, i.e., marking $M$ reached with the prefix states and the cumulative cost of its moves. The maximum memory consumption at any point is therefore $n \times p + |R_\mathcal{C}|$ states, where $p$ is the average number of states over the $n$ multi-state prefix-alignments in $D_\mathcal{C}$.

Once the number of observed cases surpasses the limit $n$, we forget prefix-alignments of some cases to accommodate the newly arrived cases. Along with reducing the memory consumption, our goal is to also reduce any loss of the conformance of the cases. Therefore, instead of na\"ively forgetting, we prudently forget the prefix-alignments of cases according to a \textit{forgetting criteria} which is explained in the following section. As discussed in the previous paragraph, while forgetting the states of a prefix-alignment ${\overline{\gamma}}$ of a case, we retain its summary as a single special state in $R_{C}$. Upon observing an orphan event, we compute its prefix-alignment ${\overline{\gamma}}^\prime$ starting from marking $M$ of its forgotten prefix ${\overline{\gamma}}$ which we retrieve from $R_{C}$. Also, we add the retained cost $\kappa_{o}$ as \textit{residual} to that incurred by the orphan event(s) so that the effective trace fitness cost of $c$ is $\kappa_{o} + \kappa({\overline{\gamma}}^\prime)$. Through the aforementioned measures, we increase the probability of correctly estimating (non)conformance of cases even with forgotten prefixes. Algorithm~\ref{alg:trace-based} provides an algorithmic summary of this proposed approach.

\begin{algorithm}[t]
\small
	\caption{Prefix-alignment-based OCC with bounded cases}\label{alg:trace-based}
	
	\begin{algorithmic}[1]

  \Require {$S \in (C \times \mathcal{A})^\ast, n$}
 	\State $i \leftarrow 0$
 \While{true}
		
			\State $i \leftarrow i+1$;
			\State $(c,a) \leftarrow S(i)$;
			\State $\overline{\gamma} \leftarrow D_{\mathcal{C}}(c,\;i-1)$;
			\If {$\overline{\gamma} \neq \emptyset$}
			\State copy all alignments of $D_{\mathcal{C}}(c^\prime,\;i-1)$ to $D_{\mathcal{C}}(c^\prime,\;i)$ for all $c^\prime \in \mathcal{C}$;
			
			\Else
			\State $\overline{\gamma} \leftarrow R_{\mathcal{C}}(c)$;
			    \If {$|D_{\mathcal{C}}| \geq n$}
			        \State select most suitable case $c^\prime \in D_{\mathcal{C}}$ through \textit{forgetting criteria};
			        
			        \State $R_{\mathcal{C}} \leftarrow \langle (\emptyset,\kappa(\overline{\gamma}_{c^\prime}), M({\overline{\gamma}}_{c^\prime}(|\overline{\gamma}_{c^\prime}|))\rangle$; 
  		  \small

			        \State forget $c^\prime$ of $D_{\mathcal{C}}$;
		        \EndIf
			\EndIf

  		\State compute $\overline{\gamma}^\prime$ through \textit{model semantics} or \textit{shortest path search}~\cite{van2019online}
  		\State $D_{\mathcal{C}}(c,\;i) \leftarrow {\overline{\gamma}}^\prime$
\EndWhile
	\end{algorithmic} 
\end{algorithm}

\paragraph{\textbf{Forgetting Criteria}:}

Our forgetting criteria consist of a set of conditions. A single-pass forgetting approach traverses through $D_\mathcal{C}$ and assigns cases residing therein a forgetting preference score in accordance with the condition it qualifies. Once a case with a certain forgetting preference is found, we narrow the search to cases with higher forgetting preferences. The only exception to the aforementioned optimization is finding a case qualifying Condition 1 below, where we completely stop the search process as we already found the most suitable case to be forgotten. In the following, we briefly explain these conditions:
\begin{enumerate}
\item The first condition looks for a complaint monuple case, i.e., case with a single event $e$, such that $\exists_{t \in T}(\lambda(t) = \#_{activity}(e) \wedge M_{i} [ \, t \rangle)$. In absence of noise, the prefix alignment of the orphan events of such a case will likely still be a global optimum. Such a case therefore is assigned the highest forgetting preference.
\item Cases with \textit{residual cost} $>0.0$ imply that their forgotten prefix was non-conformant. Since the prefix-alignment for a forgotten prefix cannot be revisited, these cases will remain non-conformant forever. Such cases are assigned with the second highest forgetting preference.
\item The prefix-alignments of complete conformant cases, i.e., cases with a 0.0 fitness cost, are optimal. The prefix-alignments of their orphan events are expected to remain optimal starting from their current marking. Such cases are therefore assigned the third highest forgetting preference.
\item Cases with zero residual cost but non-zero total fitness cost imply that their in-memory events are not fitting. We assign such cases a least forgetting preference in light of the probability that they can get optimal prefix-alignments $\overline{\gamma}_{opt}$ with less fitness cost in the future. 
\end{enumerate}

\begin{figure*}[htbp]
     \centering
     \begin{subfigure}[b]{0.40\textwidth}
         \centering
         \includegraphics[width=\textwidth]{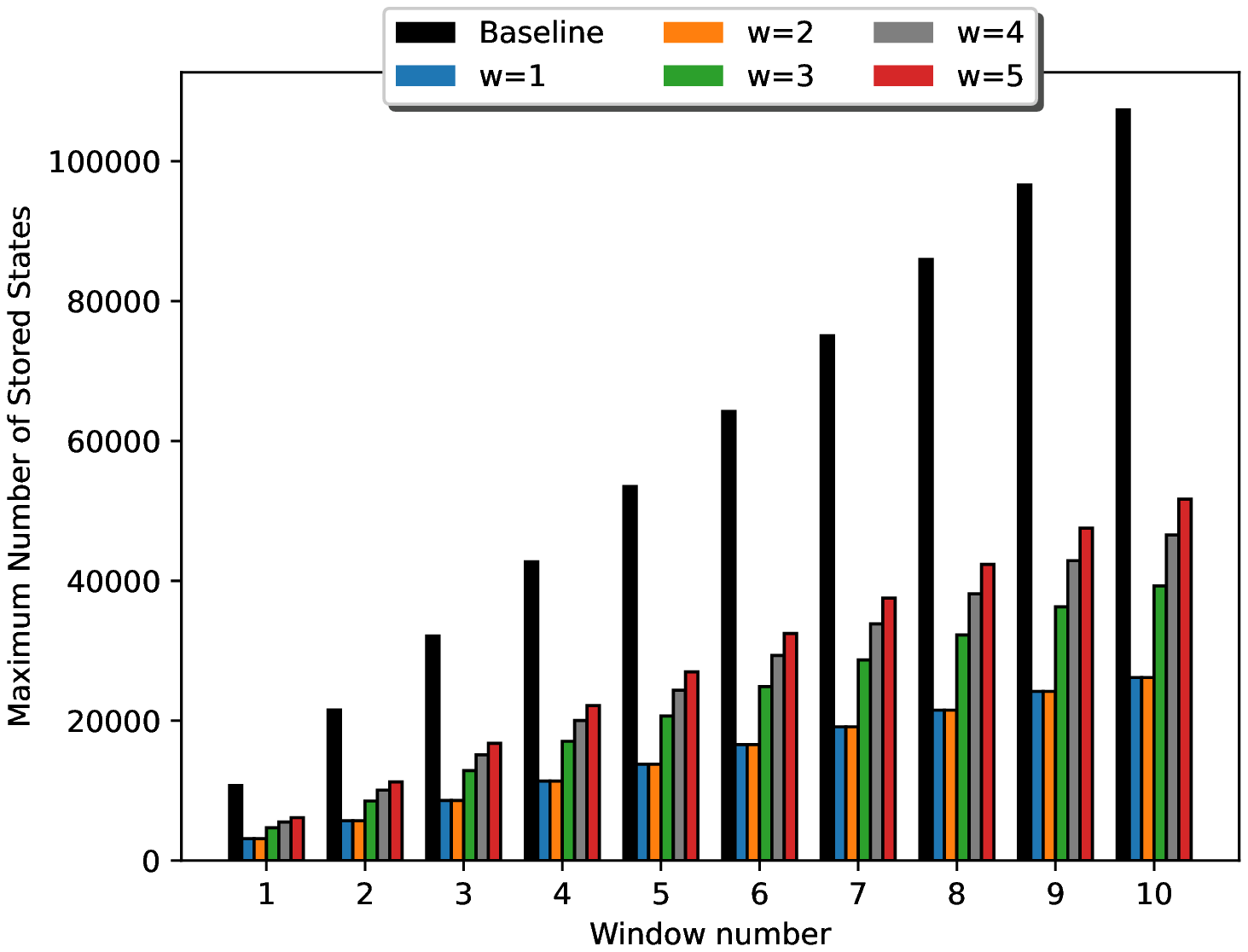}
         \caption{Stored states}
         \label{fig:exp1states}
     \end{subfigure}
     \hfill
     \begin{subfigure}[b]{0.40\textwidth}
         \centering
         \includegraphics[width=\textwidth]{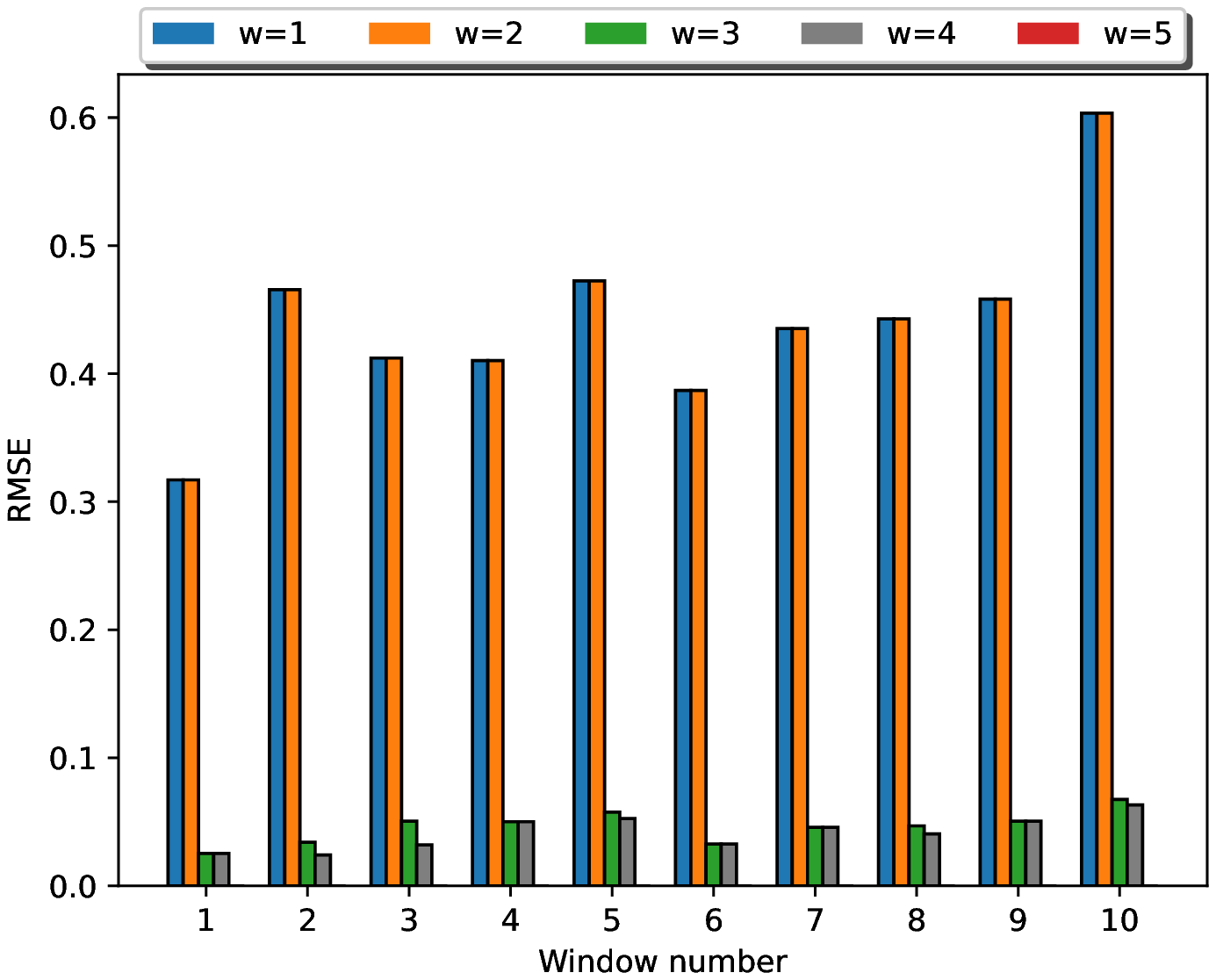}
         \caption{Fitness costs}
         \label{fig:exp1costs}
     \end{subfigure}
     \hfill
     \begin{subfigure}[b]{0.40\textwidth}
         \centering
         \includegraphics[width=\textwidth]{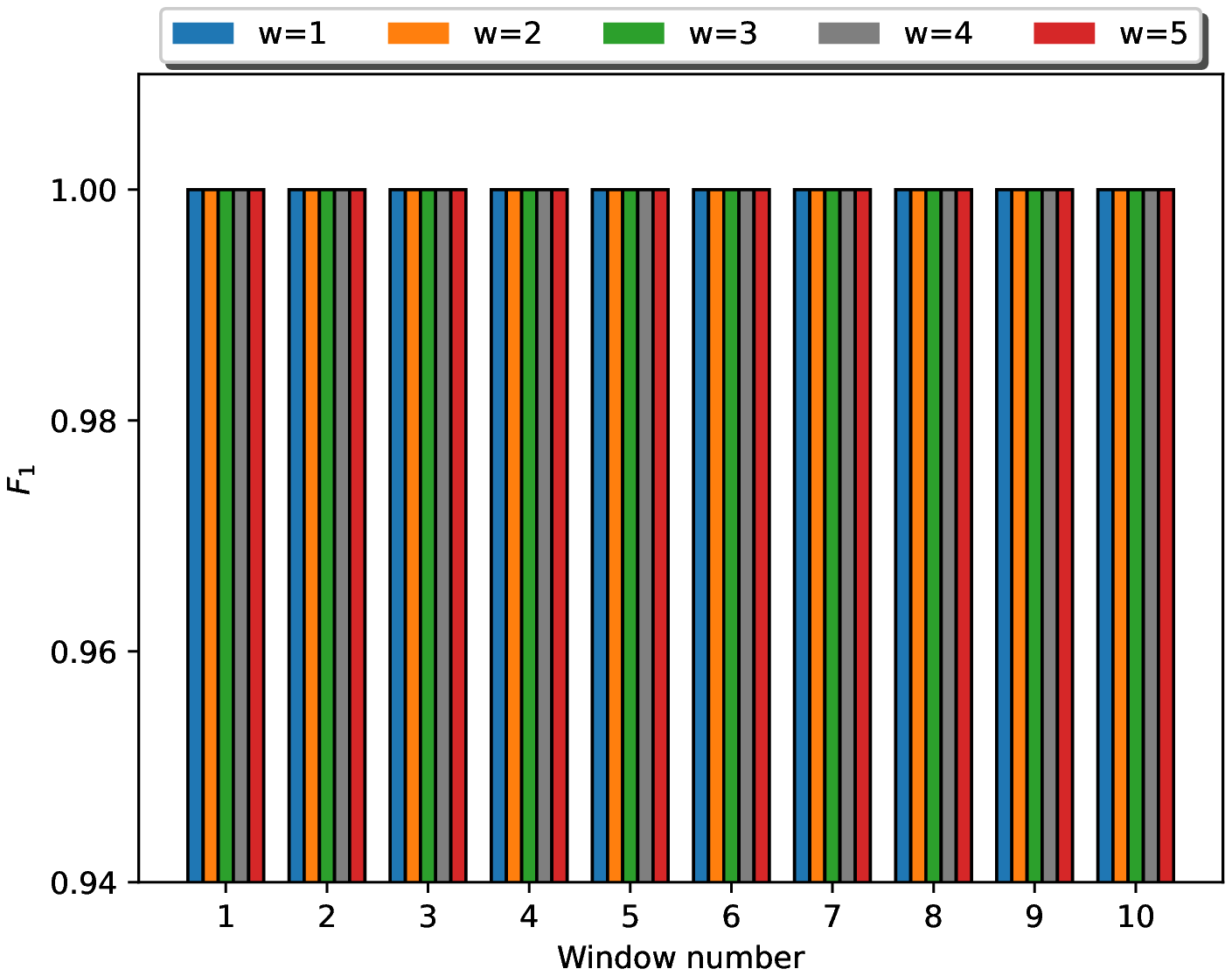}
         \caption{Classification of cases}
         \label{fig:exp1cases}
     \end{subfigure}
     \hfill
     \begin{subfigure}[b]{0.40\textwidth}
         \centering
         \includegraphics[width=\textwidth]{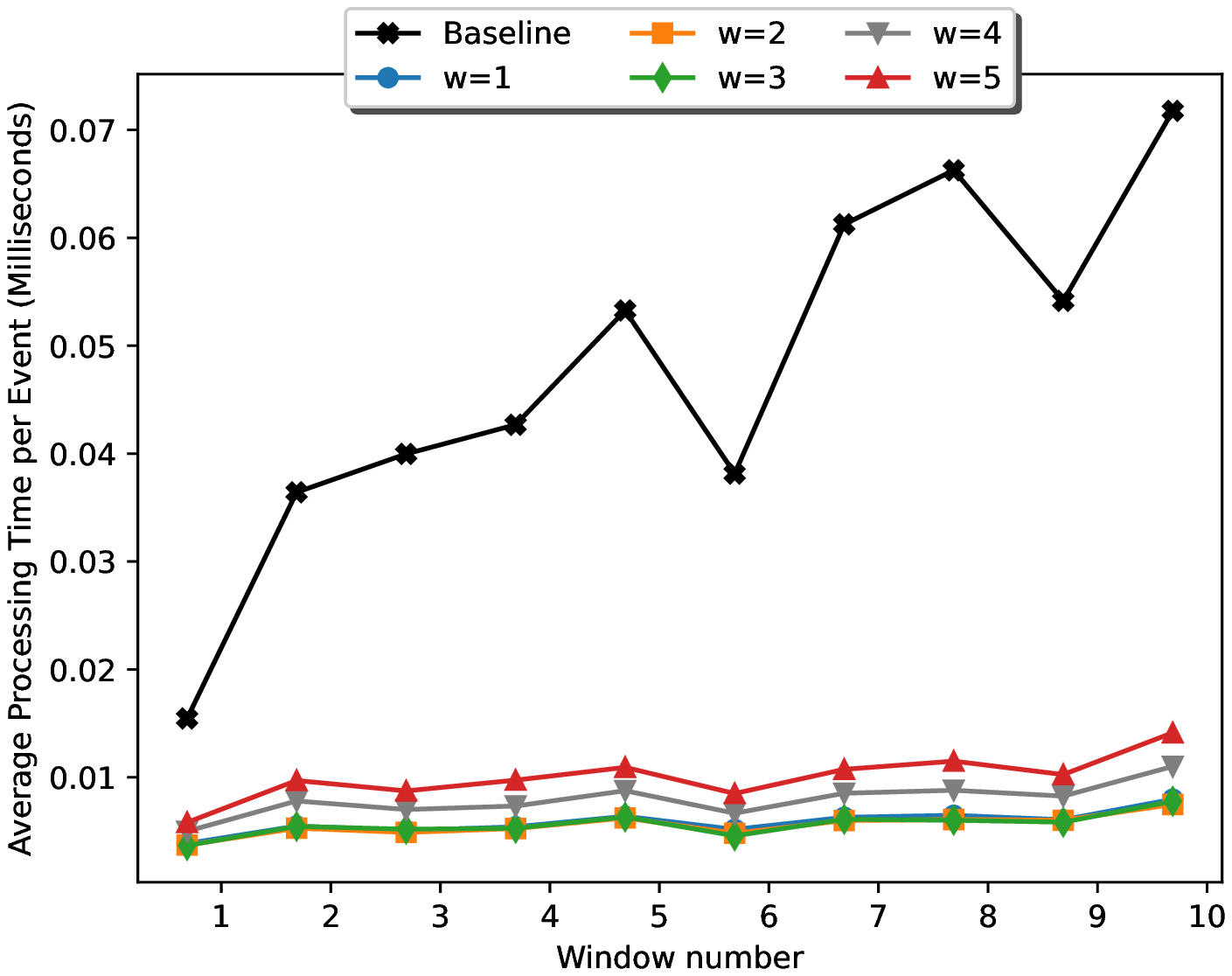}
         \caption{Average processing time per event}
         \label{fig:exp1atpe}
     \end{subfigure}
        \caption{Results of the experiment with bounded states $w$.}
        \label{fig:experiment1}
\end{figure*}

\subsection{Combined bounded cases and states with carryforward marking and cost}\label{algo:fixedcases}

The approach presented in Section~\ref{sect:fixedstates} reduces the memory consumption by limiting the number of states to $w$ which each case in $D_\mathcal{C}$ can retain. The approach presented in Section~\ref{sect:fixedcases} reduces the memory consumption by limiting $|D_\mathcal{C}|$ to $n$. Our third approach combines these two approaches by limiting the number of cases in $D_\mathcal{C}$ to $n$ and the number of states that can be retained by each of these $n$ cases to $w$. Since both $|D_\mathcal{C}|$ and the number of states per case in $D_\mathcal{C}$ is bounded, this deterministically reduces memory consumption with respect to both previously presented approaches.

As in Section~\ref{sect:fixedstates}, we forget the earliest prefix state(s) in excess of $w$ on every prefix-alignment computation for cases in $D_\mathcal{C}$ and prepend their summary as a special state to the surviving states. As in Section~\ref{sect:fixedcases}, the prefix-alignments of the cases in excess of $n$ are forgotten using the forgetting criteria and their summary is individually stored in $R_\mathcal{C}$ as a single state. Upon observing an orphan event of a case, we compute its prefix-alignment ${\overline{\gamma}}^\prime$ starting from marking $M$ of its forgotten prefix ${\overline{\gamma}}$ which we retrieve from $R_{C}$. We add the cost retained therein as \textit{residual} to the cost incurred by the orphan event(s). Hence, the effective trace fitness cost of such a case takes into account the cost of its forgotten prefix states as well. The maximum memory consumption at any point therefore is $n \times w + |R_\mathcal{C}|$ states. Algorithm~\ref{alg:combined} provides an algorithmic summary of this proposed approach.

\begin{algorithm}[t]
\small
	\caption{Prefix-alignment-based OCC with bounded cases and states}\label{alg:combined}
	
	\begin{algorithmic}[1]
 \Require {$S \in (C \times \mathcal{A})^\ast, n, w$}
 	\State $i \leftarrow 0$
 \While{true}
		
			\State $i \leftarrow i+1$;
			\State $(c,a) \leftarrow S(i)$;
			\State $\overline{\gamma} \leftarrow D_{\mathcal{C}}(c,\;i-1)$;
			\If {$\overline{\gamma} \neq \emptyset$}
			\State copy all alignments of $D_{\mathcal{C}}(c^\prime,\;i-1)$ to $D_{\mathcal{C}}(c^\prime,\;i)$ for all $c^\prime \in \mathcal{C}$;
			
			\Else
			\State $\overline{\gamma} \leftarrow R_{\mathcal{C}}(c)$;
			    \If {$|D_{\mathcal{C}}| \geq n$}
			        \State select most suitable case $c^\prime \in D_{\mathcal{C}}$ through \textit{forgetting criteria};
			        \State $R_{\mathcal{C}} \leftarrow \langle (\emptyset,\kappa(\overline{\gamma}_{c^\prime}), M({\overline{\gamma}}_{c^\prime}(|\overline{\gamma}_{c^\prime}|))\rangle$; 
  		  \small
			      
			        \State forget $c^\prime$ of $D_{\mathcal{C}}$;
		        \EndIf
			\EndIf
  		\State compute $\overline{\gamma}^\prime$ through \textit{model semantics} or \textit{shortest path search}~\cite{van2019online}
  		
  		\If {$|\overline{\gamma}^\prime| > w$}
  		\State ${\overline{\gamma}}_{o} = \langle (\emptyset,\kappa({\langle \overline{\gamma}^\prime(i) \rangle}_{i = 1 \dots |\overline{\gamma}^\prime|-w+1}), M({\overline{\gamma}^\prime}(|\overline{\gamma}^\prime|-w+1)))\rangle$ 
  		 \State $\overline{\gamma}^\prime = \overline{\gamma}_{o} \cdot {\langle \overline{\gamma}^\prime(i) \rangle}_{i = |\overline{\gamma}^\prime|-w+2 \dots |\overline{\gamma}^\prime|}$;
  	
  		  \small
  		 \EndIf
  		 \State $D_{\mathcal{C}}(c,\;i) \leftarrow \overline{\gamma}^\prime$;
\EndWhile
	\end{algorithmic} 
\end{algorithm}

\section{Experimental Evaluation}\label{sect:evaluation}
The proposed approaches are evaluated through a prototype implementation\footnote{\url{https://www.github.com/rashidzaman84/MemoryEfficientOCC}}. It is dependent on the Online Conformance package~\cite{van2019online} which uses the $A^*$ algorithm for \textit{shortest path search based prefix-alignment} computation. This package requires a Petri net process model $N$, its initial marking $M_{i}$, and its final marking $M_f$. Additionally, our first approach requires the state limit $w$, the second approach requires the case limit $n$, while the third approach requires both $w$ and $n$. The experiments are conducted on a Windows 10 64 bit machine with an Intel Core i7-7700HQ 2.80GHz CPU and 32GB of RAM.

We use the \textit{application} process and its integral \textit{offer} subprocess event data of Business Process Intelligence Challenge (BPIC’12)\footnote{\url{http://www.win.tue.nl/bpi/2012/challenge}} in all our experiments. This real event data is related to loan applications made to a Dutch financial institute and contains 13087 cases consisting of 92093 events. The reference process model has been developed by a process modelling expert in consultation with the domain knowledge experts from the financial institute. The reasons for selecting this event data include: 1) the high complexity of the reference process model, 2) the multiple types of event noise prevailing in the data, and 3) the high arrival rate of cases. We realize an event stream through dispatching events in the data on the basis of their actual timestamps. This way we ensure that the case and event distributions of the data are preserved and that also the number of cases running in parallel surpasses the limit $n$.

\begin{figure*}[h]
     \centering
     \begin{subfigure}[b]{0.40\textwidth}
         \centering
         \includegraphics[width=\textwidth]{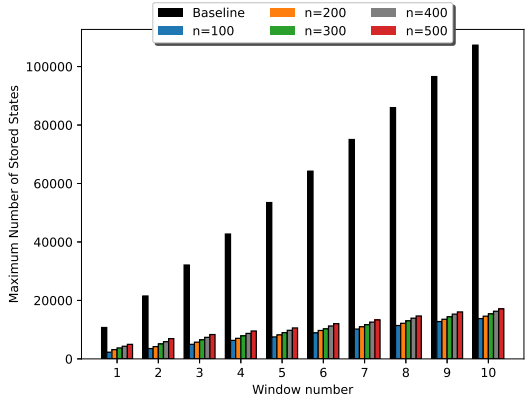}
         \caption{Stored states}
         \label{fig:exp2states}
     \end{subfigure}
     \hfill
     \begin{subfigure}[b]{0.40\textwidth}
         \centering
         \includegraphics[width=\textwidth]{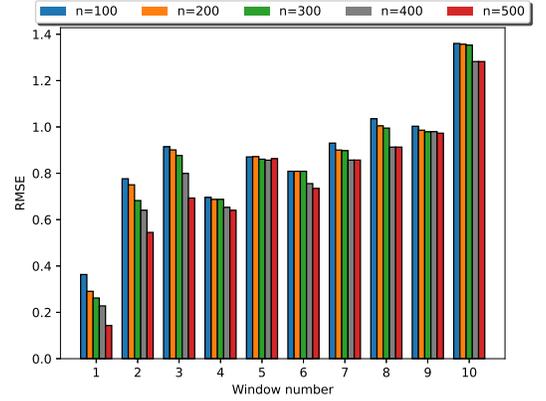}
         \caption{Fitness costs}
         \label{fig:exp2costs}
     \end{subfigure}
     \hfill
     \begin{subfigure}[b]{0.40\textwidth}
         \centering
         \includegraphics[width=\textwidth]{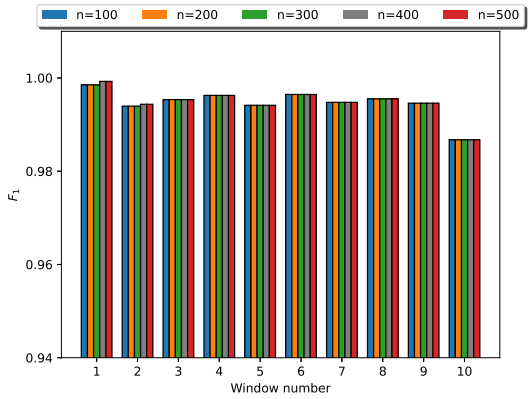}
         \caption{Classification of cases}
         \label{fig:exp2cases}
     \end{subfigure}
     \hfill
     \begin{subfigure}[b]{0.40\textwidth}
         \centering
         \includegraphics[width=\textwidth]{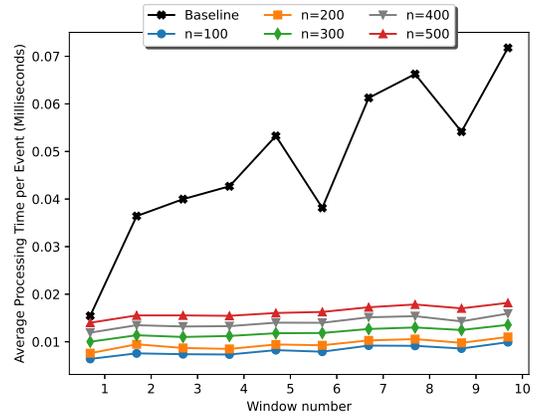}
         \caption{Average processing time per event}
         \label{fig:exp2atpe}
     \end{subfigure}
        \caption{Results of the experiment with bounded cases $n$.}
        \label{fig:experiment2}
\end{figure*}

\begin{figure*}[h]
     \centering
     \begin{subfigure}[b]{0.40\textwidth}
         \centering
         \includegraphics[width=\textwidth]{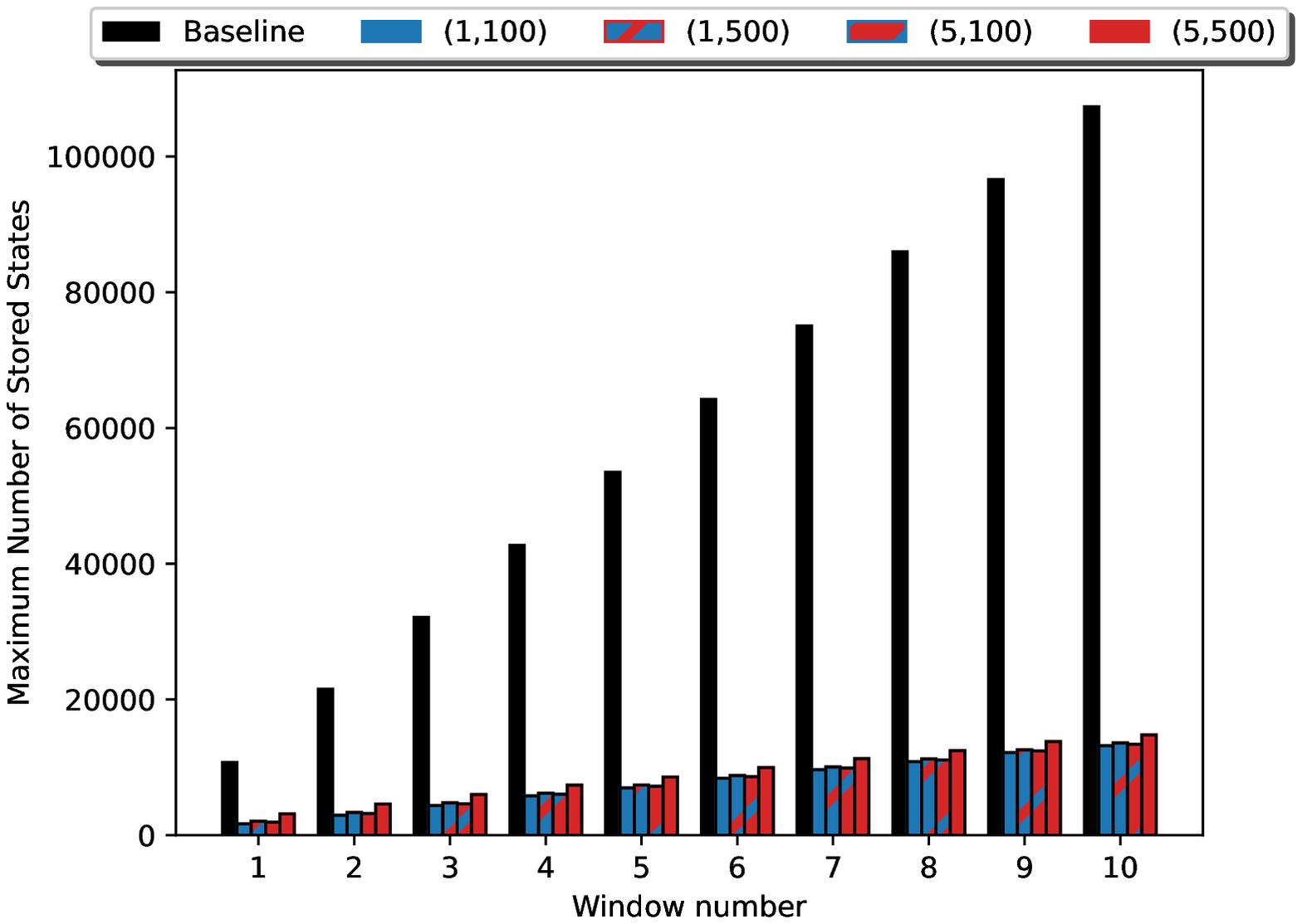}
         \caption{Stored states}
         \label{fig:exp3states}
     \end{subfigure}
     \hfill
     \begin{subfigure}[b]{0.40\textwidth}
         \centering
         \includegraphics[width=\textwidth]{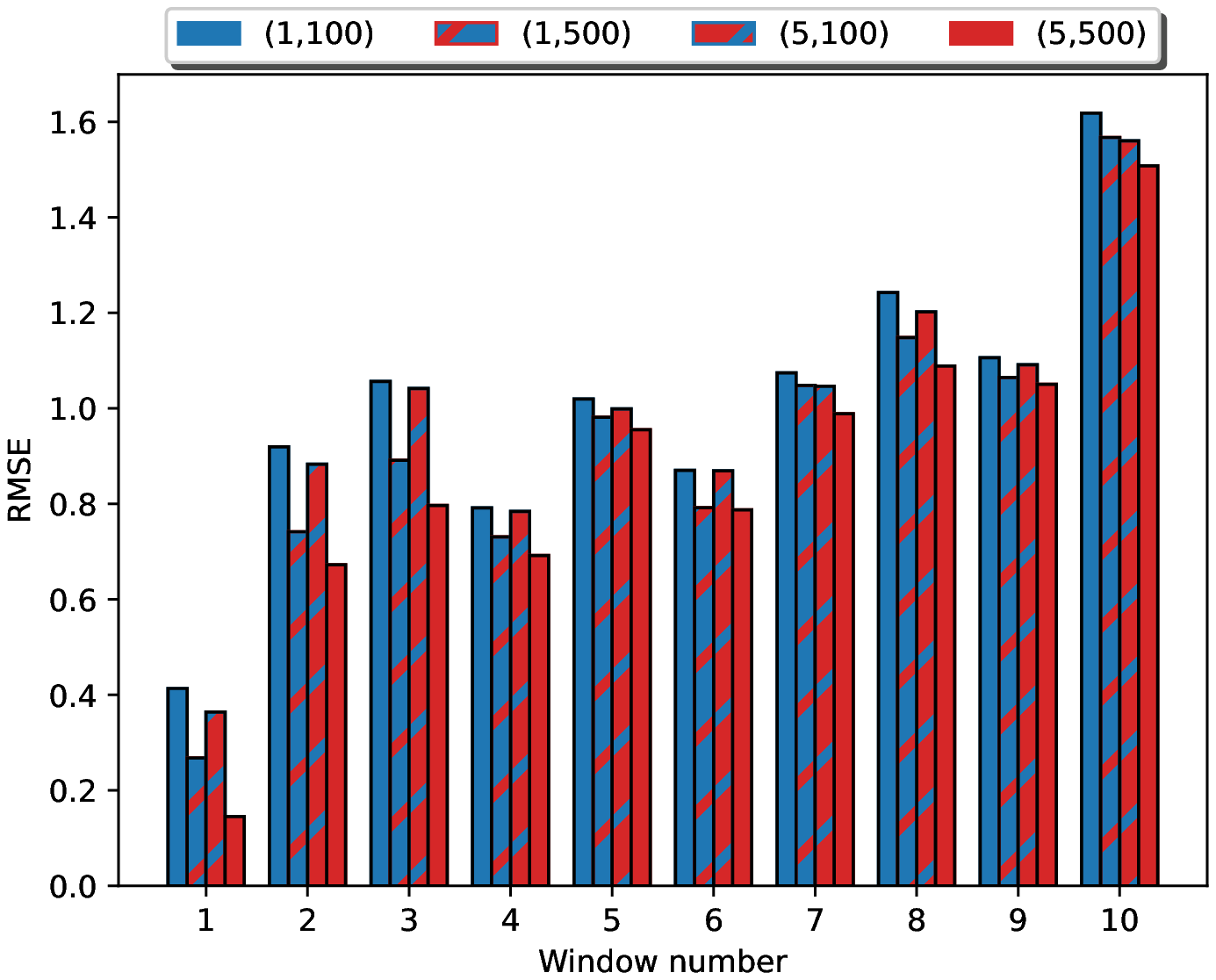}
         \caption{Fitness costs}
         \label{fig:exp3costs}
     \end{subfigure}
     \hfill
     \begin{subfigure}[b]{0.40\textwidth}
         \centering
         \includegraphics[width=\textwidth]{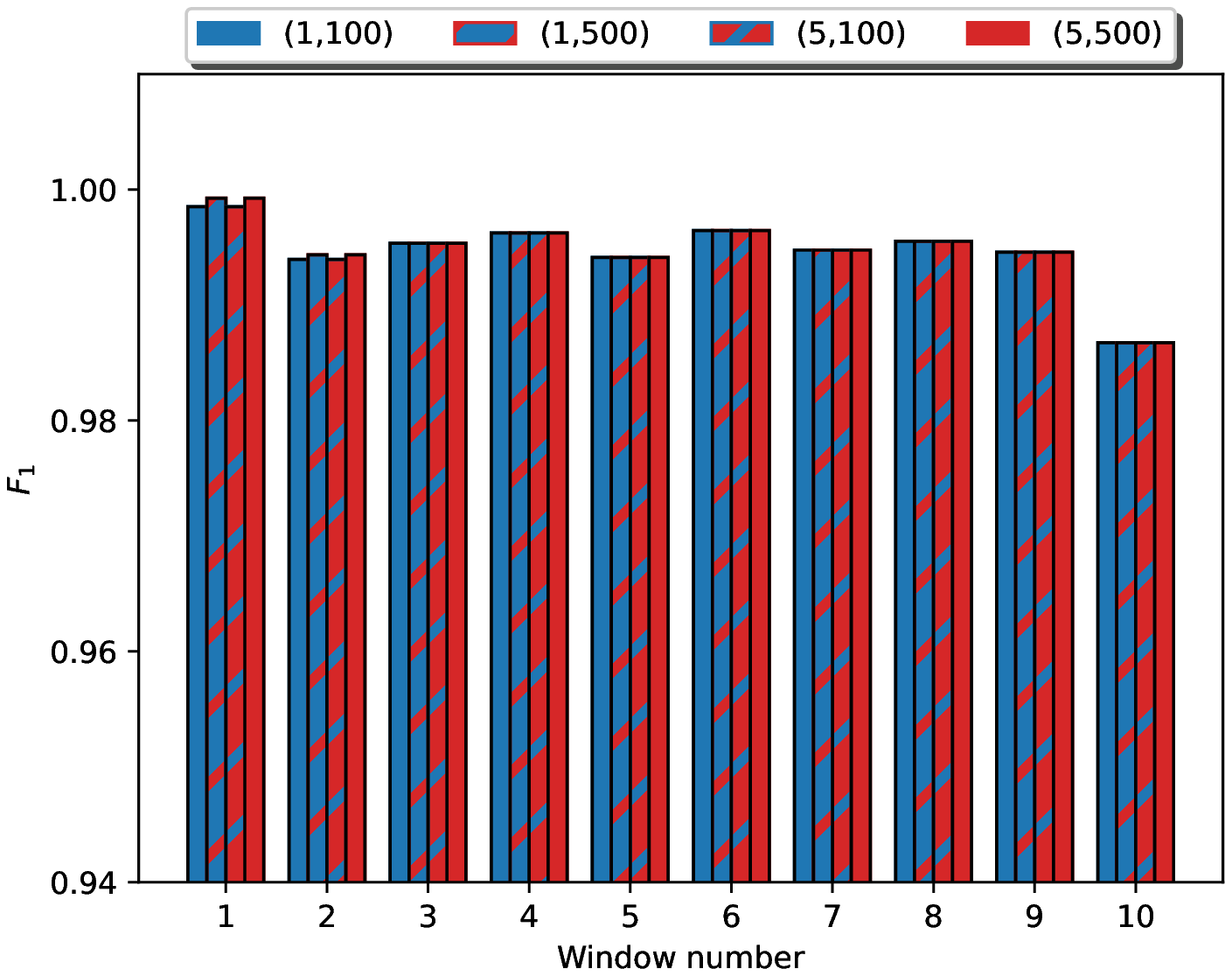}
         \caption{Classification of cases}
         \label{fig:exp3cases}
     \end{subfigure}
     \hfill
     \begin{subfigure}[b]{0.40\textwidth}
         \centering
         \includegraphics[width=\textwidth]{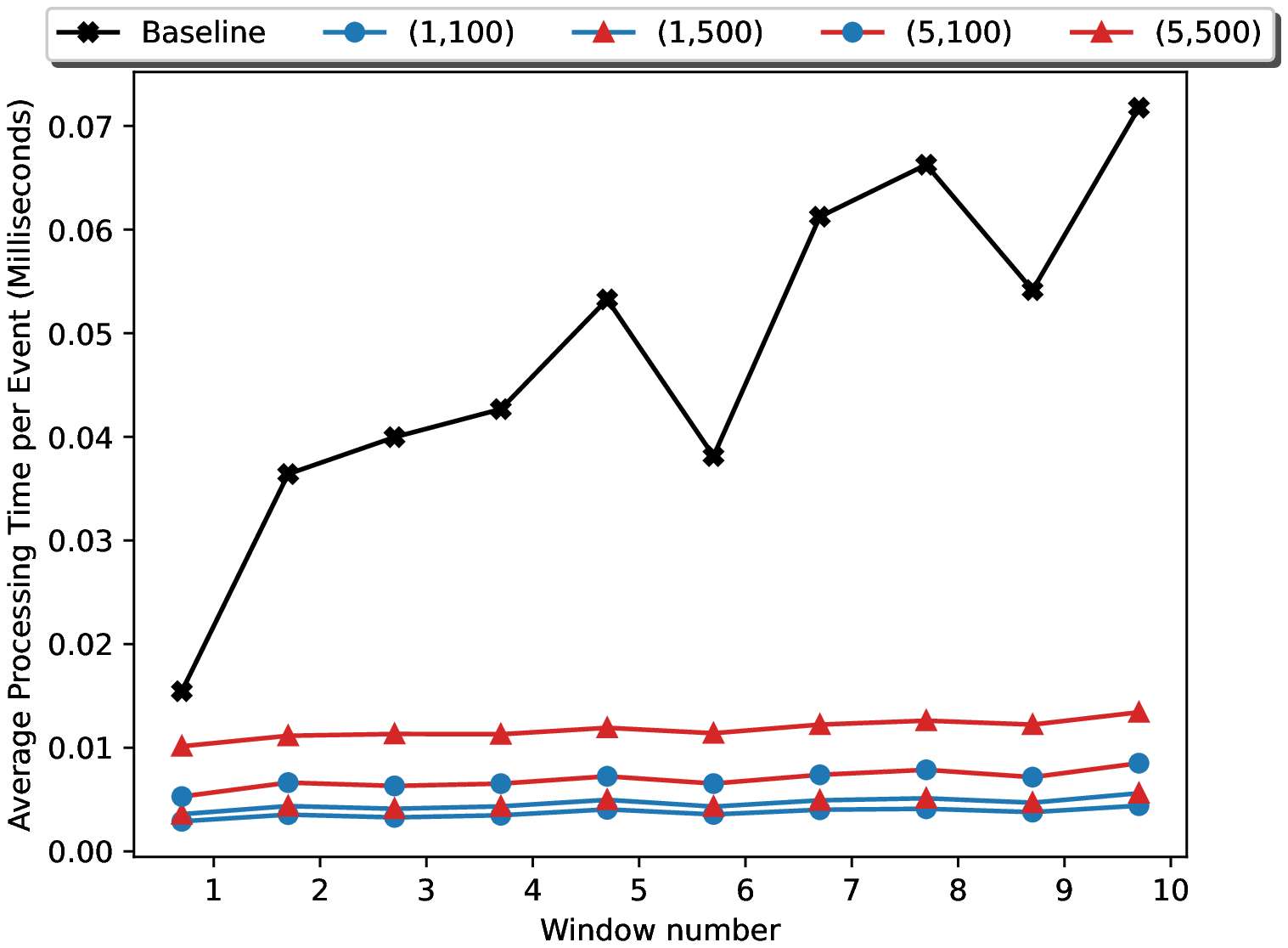}
         \caption{Average processing time per event}
         \label{fig:exp3atpe}
     \end{subfigure}
        \caption{Results of the experiment with bounded cases $n$ and states $w$, indicated as $(w,n)$}
        \label{fig:experiment3}
\end{figure*}

We are presenting the results in an event-window notion where all the windows consist of the same number of observed events. We are considering state of the art incremental prefix-alignments with an
infinite memory, hence always resulting in optimal prefix-alignments, as our \textit{baseline}. For each event-window, the results of our experiments are comprised of four statistics: the number of maximum states in memory (representing memory footprint), the root mean square error (RMSE) of the fitness cost, the $F_1$ for classification of cases as conformant or non-conformant, and the average processing time per event (APTE). To calculate the \emph{trace fitness cost}, all our experiments use the default unit cost of $1.0$ for both log and model moves, while synchronous and model moves with silent transition $\tau$ incur a cost of $0.0$. The RMSE and $F_1$ are calculated with reference to the baseline. We use five state limits, i.e., $w$=1, 2, 3, 4, and 5, in experiments with the first approach. For experiments with our second approach, we use case limits $n$= 100, 200, 300, 400, and 500. To evaluate our third approach, we use combinations of state limits $w$= 1, 2, $\dots$, 5 with case limits $n$= 100, 200, $\dots$, 500. We consider both $w$ and $n$ equal to $\infty$, i.e., infinite memory, for our baseline. Note that the Y-axis of the $F_1$ does not start with 0.0 but starts with 0.94 for highlighting the minor differences. For APTE, we replicate the event stream $50$ times, renaming the cases in each replication. This effectively means we process $50 \times 13087$ cases and $50 \times 90923$ events, mimicking a much larger event stream. We report the APTE as the mean value over these $50$ iterations.

\subsection{Results}

Our first set of experiments, the results of which are provided in Figure~\ref{fig:experiment1}, are related to the bounded states related approach presented in Section~\ref{sect:fixedstates}. As expected and depicted in Figure~\ref{fig:exp1states}, we store fewer cumulative states for all the state limits $w$=1, 2, 3, 4, and 5 in comparison to the baseline. The states storage is similar for states limit of 1 and 2 as for the former we need to retain the special summary state in addition to the allowed one state. The results for fitness costs, as depicted in Figure~\ref{fig:exp1costs}, are quite interesting. Even with retaining very few states per case, the RMSE is not significant. With being close to zero for $w$=3 and 4, it is exactly zero for $w$=5. Referring to Figure~\ref{fig:exp1cases}, the statistics regarding the classification of cases is also interesting. For all our state limits $w$, the $F_1$ of 1.0 indicates that our approach always correctly classifies cases as either conformant or non-conformant.

Referring to Figure~\ref{fig:exp1atpe}, all the state limits $w$ are performing far better and almost persistent on the APTE metric as well. Interestingly, the curve for baseline in Figure~\ref{fig:exp1atpe} is depicting an increasing trend portraying that the APTE is increasing with increase in observed (and accordingly stored) events and cases which is actually misleading. Through further experiments and analysis, we realized that the increasing trend is mainly attributed to 1) the uneven number of shortest path search based prefix-alignment computations among the event-windows and 2) the trace-length directly contributing to the computational complexity of $A^*$ algorithm. The proposed approach is therefore stream-friendly in terms of computation performance.

Our second set of experiments, results provided as Figure~\ref{fig:experiment2}, are related to the bounded cases related approach presented in Section~\ref{sect:fixedcases}. As depicted in Figure~\ref{fig:exp2states}, our proposed approach is highly frugal to storing states in all the case limits $n$ in comparison to the baseline. However, the differences between these varying case limits are not significant. In all these case limits, prefix alignments are continuously forgotten to remain in the limit $n$, therefore cases do not grow significantly in terms of the number of states in their prefix-alignments. Hence, the maximum number of states retained in all these case limits is somehow comparable.

Referring to Figure~\ref{fig:exp2costs}, we observe a higher RMSE with respect to the previous approach, implying that this approach generally overestimates the fitness cost of cases. Although not proportionally, the RMSE decreases with increase in the case limit $n$. Investigation of this overestimation revealed an interesting factor. There are two completely identical execution sequences in the process model which lead to two different reachable markings. This anomalous behavior exists because two transitions share the same label and input place and hence are in a choice relation. The incremental prefix-alignments approach, lacking any information about the future events, always fires the transition with which the final marking can be reached earlier. A fraction of the future events of some of the cases corresponds to the transitions belonging to the execution sequence of the alternate transition. For such cases, the alternate transition should have been fired instead. A considerable number of cases are forgotten at this stage by our forgetting approach. Since the marking $M$ of the forgotten states is considered as an initial marking $M_{i}$ for computing the prefix-alignments for their orphan events, the mentioned fraction of orphan events does not fit anymore and hence is improperly considered as log moves. Linked to the aforementioned effect, some events prematurely lead their cases to the final marking and hence all the preceding events are improperly marked as log moves.

Referring to Figure~\ref{fig:exp1cases}, the $F_1$ for all the $n$ values is almost 1.0. Hence, besides the overestimation of non-conformance discussed in the previous paragraph, this approach is highly accurate in binary classification, i.e., conformant and non-conformant cases. Regarding the APTE, due to the continuous forgetting of cases, the state size of multi-state $n$ cases does not grow significantly and hence even the shortest path search based prefix-alignment computations do not take long. Therefore, all the case limits $n$ sustain almost a uniform APTE, as can be seen in Figure~\ref{fig:exp2atpe}. Hence, this approach is also computation-wise stream-friendly.

Our third set of experiments, results provided in Figure~\ref{fig:experiment3}, are related to the approach combining bounded cases and bounded states which is presented in Section~\ref{sect:fixedcases}. For clarity, we are reporting the results for only $w$= 1 and 5 in combination with $n$=100 and 500. As evident in Figure~\ref{fig:exp3states}, we are highly frugal to consuming states in the reported state and case limit combinations. Interestingly, the state consumption is comparable for the same state limit even with different case limits. This peculiarity can be explained by the fact that in all the $n$ limits the prefix-alignments are frequently forgotten such that the multi-state cases do not necessarily reach the limit $w$. Referring to Figure~\ref{fig:exp3costs}, as expected, the two-dimensional bounding causes a slight increase in RMSE with respect to the previous experiments. We can notice that increasing the $w$ and $n$ (disproportionately) reduces the RMSE. Interestingly, we observe an $F_1$ close to 1.0 for all the $w$ and $n$ combinations in Figure~\ref{fig:exp3cases}. Regarding APTE, referring to Figure~\ref{fig:exp3atpe}, as with the previous experiments, this approach also performs far better and persistently in all $w$ and $n$ combinations.

\subsection{Discussion}

We stressed our proposed approaches with a high noise-bearing event data where only 17\% of the cases having a prefix-length of 9 or more are conformant. 45\% cases of the 83\% non-conformant cases have multiple events with, probably different types of, associated noise. The multiple identical execution sequences leading to different markings property of the reference process model added interesting dimensions to the experiments. We conclude that the bounded states approach with a suitable state limit is very light on memory and performs as good as the baseline. A suitable state limit mainly depends on the number of past events required to revisit a prefix-alignment for optimality. For the approach with bounded cases, though we save considerably on the memory, the fitness costs are overestimated. A case limit equal to the number of maximum in-parallel running non-conformant cases will result in statistics that are as good as those of the baseline, which can likely be the situation in real event streams. The combined bounded cases and bounded states approach is much more suitable for processes with long traces and a binary classification task of cases as conformant or non-conformant. However, all these presented approaches may at some point grow unbounded, with the second and third approaches relatively withstanding for long as the states summary in $R_{C}$ consists of just a single state. Therefore, a systematic mechanism to completely forget cases should complement these approaches. Besides the problem arising due to anomalous execution sequences in process models, log moves at exactly the end of a prefix-alignment may get unaccounted in the future if the prefix-alignment is forgotten at this stage, even with retaining the current marking $M$.

\section{Conclusion and Future Work}\label{sect:conclusions}
We presented three incremental approaches to deal with scarcity of memory and at the same time avoid missing-prefix problem in prefix-alignment-based OCC of event streams. The effectiveness of these approaches is established through experiments with real-life event data by mimicking it as an event stream. The proposed approaches considerably reduce memory consumption and positively impact the overall processing time of events. These approaches are equally applicable in general (prefix)alignments and can easily be extended to other CC techniques.

Based on the observations and findings of the conducted experiments, we foresee the need for devising techniques to completely bound memory in event streams processing applications. Our last two proposed approaches suffer from the `hard-coding of initial marking' problem caused by anomalous execution sequences existing in process models. A technique that, upon reaching a threshold of non-conformance in orphan events, revisits the past alignment decisions can mitigate the mentioned problem.

\begin{acks}
 The authors have received funding within the BPR4GDPR project from the European Union’s Horizon 2020 research and innovation programme under grant agreement No. 787149.

\end{acks}

\bibliographystyle{ACM-Reference-Format}
\bibliography{SAC2022} 

\end{document}